\documentclass[twocolumn,superscriptaddress,showpacs,nofootinbib,preprintnumbers,secnumarabic,amssymb, nobibnotes, aps, prd]{revtex4-2}
\usepackage[utf8]{inputenc}
\usepackage{graphicx}
\usepackage{latexsym,amsmath,amssymb,amsthm,lmodern,float,url}
\usepackage{natbib}
\usepackage{color}
\usepackage{microtype}
\usepackage{import}
\usepackage{bbold}
\usepackage[plain]{fancyref}
\usepackage{varioref}
\usepackage{slashed}
\usepackage{multirow}
\usepackage{tikz}
\usepackage{scrextend}
\usepackage{braket}
\usetikzlibrary{shapes}
\usetikzlibrary{positioning}
\usepackage[normalem]{ulem}

\newcommand{\fig}[1]{Fig.~\ref{fig:#1}}
\newcommand{\tab}[1]{Tab.~\ref{tab:#1}}
\newcommand{\eq}[1]{Eq.~(\ref{eq:#1})}

\usepackage[colorlinks=true,backref=false, linktocpage=true,
citecolor=blue,urlcolor=blue,linkcolor=blue,pdfpagemode=UseOutlines]{hyperref}
\hypersetup{%
  bookmarksnumbered=true,
  pdftitle = {},
  pdfsubject = {},
  pdfauthor = {},
  pdfkeywords = {}
}

\DeclareMathOperator{\tr}{Tr}
\DeclareMathOperator{\re}{Re}

\newcommand{\bi}{\mathbb{BI}}

\begin{document}
\preprint{FERMILAB-PUB-21-702-T}
\title{Gauge Theory Couplings on Anisotropic Lattices}
\author{Marcela Carena}
\email{carena@fnal.gov}
\affiliation{Fermi National Accelerator Laboratory, Batavia,  Illinois, 60510, USA}
\affiliation{Enrico Fermi Institute, University of Chicago, Chicago, Illinois, 60637, USA}
\affiliation{Kavli Institute for Cosmological Physics, University of Chicago, Chicago, Illinois, 60637, USA}
\affiliation{Department of Physics, University of Chicago, Chicago, Illinois, 60637, USA}
\author{Erik J. Gustafson}
\email{egustafs@fnal.gov}
\affiliation{Fermi National Accelerator Laboratory, Batavia,  Illinois, 60510, USA}
\author{Henry Lamm}
\email{hlamm@fnal.gov}
\affiliation{Fermi National Accelerator Laboratory, Batavia,  Illinois, 60510, USA}
\author{Ying-Ying Li}
\email[Corresponding author: ]{yingying@fnal.gov}
\affiliation{Fermi National Accelerator Laboratory, Batavia,  Illinois, 60510, USA}
\author{Wanqiang Liu}
\email{wanqiangl@uchicago.edu}
\affiliation{Department of Physics, University of Chicago, Chicago, Illinois, 60637, USA}

\date{\today}

\begin{abstract}
The advantage of simulating lattice field theory with quantum computers is hamstrung by the limited resources that induce large errors from finite volume and sizable lattice spacings. Previous work has shown how classical simulations near the Hamiltonian limit can be used for setting the lattice spacings in real-time through analytical continuation, thereby reducing errors in quantum simulations. In this work, we derive perturbative relations between bare and renormalized quantities in Euclidean spacetime at any anisotropy factor -- the ratio of spatial to temporal lattice spacings -- and in any spatial dimension for $U(N)$ and $SU(N)$. This  reduces the required classical preprocessing for quantum simulations. We find less than $10\%$ discrepancy between our perturbative results and those from existing nonperturbative determinations of the anisotropy for $SU(2)$ and $U(1)$ gauge theories. For the discrete groups $\mathbb{Z}_{10}$, $\mathbb{Z}_{100}$ and $\mathbb{BI}$, we perform lattice Monte Carlo simulations to extract anisotropy factors and observe similar agreement with our perturbative results.
\end{abstract}

\maketitle

\section{\label{sec:introduction}Introduction}
Quantum computers can make predictions of nonperturbative quantum field theories beyond the reach of classical resources~\cite{Feynman:1981tf,doi:10.1126/science.273.5278.1073,Jordan:2017lea} such as neutron star equations of state and the out-of-equilibrium dynamics of the early universe~\cite{Davoudi:2022cah}. 
However, quantum simulations are constrained by limited and noisy resources, and will continue to be so for the foreseeable future. Current estimates suggest $\sim 10$ logical qubits per gluon link should suffice to digitize $SU(3)$~\cite{Raychowdhury:2018osk,Raychowdhury:2019iki,Alexandru:2019nsa,Davoudi:2020yln,Ciavarella:2021nmj,Kan:2021xfc,Alexandru:2021jpm}, with similar requirements for $U(1)$ and $SU(2)$~\cite{Zohar:2012xf,Zohar:2013zla,Zohar:2014qma,Zohar:2016iic,Bender:2018rdp,Haase:2020kaj,Bauer:2021gek,Alexandru:2019nsa,Raychowdhury:2018osk,Raychowdhury:2019iki,Davoudi:2020yln,Ciavarella:2021nmj,Ji:2020kjk,PhysRevD.99.114507,Bazavov:2015kka,Zhang:2018ufj,Unmuth-Yockey:2018ugm,Unmuth-Yockey:2018xak,Wiese:2014rla,Luo:2019vmi,Brower:2020huh,Mathis:2020fuo,Liu:2021tef,Gustafson:2021qbt,Gustafson:2022xlj}. The total number of qubits required is dependent upon the phenomena one wishes to study; for a $3+1d$ lattice gauge theories, $\mathcal{O}((L/a)^3)$ links are usually required, exacerbating the qubit requirement. Due to quantum noise, quantum error correction is likely required, with an overhead of $\mathcal{O}(10^{1-5})$~\cite{ionq_2020,ibm_2021, google_2020} physical qubits per logical qubit depending on platform. Preliminary work in specialized error correction for lattice gauge theories may be able to reduce these costs~\cite{Rajput:2021trn,Klco2021Hierarchy}.

Resource estimates must also consider the gate costs to implement time evolution of the theory under a lattice Hamiltonian. 
The time evolution operator $\mathcal U(t)$ - a generically dense matrix - must usually be implemented approximately. 
Most studies of gauge theories consider the Kogut-Susskind Hamiltonian $\hat H_{KS}$~\cite{PhysRevD.11.395} together with Trotterization. Currently, $\mathcal{O}(10^{49})$ T gates are estimated to be required to compute the shear viscosity with $\mathcal{O}(10^5)$ logical qubits~\cite{Kan:2021xfc}. This upper bound can be reduced, for example by controlling only errors on low-lying states~\cite{Sahinoglu2020hamiltonian,Hatomura:2022yga} or by requiring the algorithmic error be comparable to the $\mathcal{O}(1)$ theoretical uncertainties. Even with these reductions the quantum resources are far beyond near-term devices.  Further, this estimate neglects state preparation, which often dominates the total gate costs~\cite{Jordan:2011ne}.

Given the large resource estimates, it is informative to consider the history of prime factorization~\cite{shor1994algorithms, PhysRevLett.76.3228, Zalka2006, Fowler2012,gidney2021factor} and quantum chemistry~\cite{doi:10.1126/science.1113479,doi:10.1073/pnas.1619152114,PhysRevLett.123.070503,doi:10.1073/pnas.1619152114,Lee:2020egw} where resources were reduced by using more clever quantum subroutines and performing better classical processing. Gate reductions may be possible via other approximations of $\mathcal U(t)$~\cite{PhysRevLett.123.070503,cirstoiu2020variational,gibbs2021longtime,yao2020adaptive,PhysRevLett.114.090502,Low2019hamiltonian}. Using other arithmetic subroutines can yield drastically fewer resources ~\cite{hadfield2016scientific,haner2018optimizing,Takahashi10,gidney2021factor}. Lattice-field-theory specific error correction~\cite{Rajput:2021trn,Klco2021Hierarchy} or mitigation~\cite{Stannigel:2013zka,Stryker:2018efp,Halimeh:2019svu,Lamm:2020jwv,Tran:2020azk,Halimeh:2020ecg,Halimeh:2020kyu,Halimeh:2020djb,Halimeh:2020xfd,VanDamme:2020rur,Kasper:2020owz,Halimeh:2021vzf} could help further. Recently, quantum circuits for Hamiltonians with reduced lattice artifacts~\cite{Luo:1998dx,Carlsson:2001wp} were constructed~\cite{Carena:2022kpg}. 

A full accounting of resources should take advantage of any reductions through classical computations. In quantum chemistry, improved basis functions were found which render the Hamiltonian more amenable to quantum circuits~\cite{Lee:2020egw} at the cost of classical preprocessing. In the same way, digitization in gauge theories seeks to find efficient basis states. Here, Euclidean lattice simulations on classical computers help quantify the scheme-dependent systematic errors~\cite{Hackett:2018cel,Alexandru:2019nsa,Ji:2020kjk,Ji:2022qvr,Alexandru:2021jpm,Hartung:2022hoz}. We can draw another analogy to the case of prime factorization where Eker{\aa} and H{\aa}stad's modifications ~\cite{ekeraa2016modifying,ekeraa2017quantum,ekeraa2017pp,ekeraa2018general} of Shor's algorithms~\cite{shor1994algorithms} used classical processing to reduce qubits and quantum arithmetic while increasing the success rate.  In the same way, lattice calculations have a number of steps that can potentially be offloaded to classical resources. The first suggested was to use Euclidean lattice ensembles to perform stochastic state preparation yielding shallower individual circuits~\cite{Lamm:2018siq,Harmalkar:2020mpd,Gustafson:2020yfe,Yang:2021tbp}. Further, classical simulations can be used to set the scales, which via analytical continuation~\cite{Osterwalder:1973dx,Osterwalder:1974tc} gives the lattice spacings of the quantum simulation with few or no quantum resources ~\cite{Carena:2021ltu,Clemente:2022cka}.  

Although in~\cite{Carena:2021ltu} the connection between lattice Hamiltonian at finite real-time temporal lattice spacing $a_t$ and Euclidean temporal lattice spacing $a_\tau$ was made, the final step of connecting the Hamiltonian to the bare parameters used in Euclidean action was missing. The brute-force approach would compute multiple anisotropies $\xi = a/a_\tau$ for a fixed spatial lattice spacing $a$ and then extrapolate to the desired $\xi$ used in the quantum simulation. This is analogous to studies of the relation between Euclidean and Hamiltonian limits~\cite{Hamer:1995zj,Byrnes:2003gg}. Continuum extrapolations requires simulations at multiple lattice spacings. While this will become the practice as quantum lattice simulations becomes a precision endeavor, for now quantum noise and low shot rates dominate the error budget of calculations~\cite{Ciavarella:2021lel,Farrell:2022wyt,Rahman:2022rlg}, burying errors from determining $\xi$. Thus, the idea of a perturbative calculations of $\xi$ becomes attractive, as it directly gives a fixed $\xi$ trajectory in terms of the bare parameters. This implies that the only the measurement of the spatial lattice spacing $a$ through Euclidean simulation is required -- reducing the classical computing resources. 
Through analytical continuation to Minkowski spacetime, spatial (temporal) spacings, $a$ ($a_t$), are determined for quantum simulations \cite{Carena:2021ltu}.

In this paper, we perform the one-loop perturbative calculation of $\xi$ using the background field method~\cite{DeWitt:1967ub,DeWitt:1967uc,Honerkamp:1972fd,tHooft:1973bhk}. In the early days of lattice QCD, this technique along with other methods~\cite{Callan:1978bm,Hasenfratz:1980kn} were used to compute the scale parameter $\Lambda$~\cite{Dashen:1980vm}. Of relevance to quantum simulations, this included matching isotropic $3+1d$ $SU(N)$ lattice results to the Hamiltonian limit ($\xi\rightarrow\infty$)~\cite{Hasenfratz:1981tw}. Later, this was extended to arbitrary anisotropy~\cite{Karsch:1982ve} and to the Hamiltonian limit in $2+1d$~\cite{Hamer:1996ub}. Here, we present a unified derivation of $\xi$ for $U(N)$ and $SU(N)$ for arbitrary dimensions and anisotropy.
Here we focus on the Wilson action and consider its connection to the Kogut-Susskind Hamiltonian. Similar studies can be carried out for quantum simulations of
improved Hamiltonians~\cite{Luo:1998dx,Carlsson:2001wp,Carena:2022kpg} following initial work in 3+1$d$ $SU(N)$~\cite{Iwasaki:1983zm,GarciaPerez:1996ft,Sakai:2000jm,Sakai:2003va,Drummond:2002yg}. Since continuous gauge theories can be digitized for quantum simulations with discrete subgroups, we further explore whether our perturbative calculations for the continuous group can predict discrete subgroup results.

This paper is organized as follows. In Sec.~\ref{sec:gencon}, we review the background field method and show how to perturbatively compute the renormalized anisotropy. This is followed by Sec.~\ref{sec:u1} and Sec.~\ref{sec:sun} where the special cases of $U(1)$ and $SU(N)$ respectively are investigated. We extend the calculations to $U(N)$ in Sec.~\ref{sec:un}. The anisotropy factors computed perturbatively are compared with Monte Carlo results for continuous and discrete groups in Sec.~\ref{sec:discretegroup}, to demonstrate the effectiveness of our perturbative computations. We leave Sec.~\ref{sec:Conclusions} to conclude and discuss future work. Details about the integrals involved in the perturbative calculations are in the Appendices.

\section{Background Field Method}\label{sec:gencon}
Euclidean anisotropic lattices are characterized by the anisotropy $\xi = a/a_\tau$. Throughout this work, we will use Greek indices $(\mu,\nu)$ to indicate spacetime dimensions, and Latin indices ($i,j$) to indication spatial dimensions. Consider the anistropic Wilson action:
\begin{eqnarray}
S(U) = \sum_x\bigg[ \beta_\sigma \sum_{i>j} \re\tr P_{ij} +\beta_\tau \sum_{i} \re\tr P_{0i}\bigg],
\label{eq:action}
\end{eqnarray}
with the plaquette term:
\begin{eqnarray}
P_{\mu\nu} =
\mathbb{1} - U_{x, x+\mu}U_{x+\mu, x+\mu+\nu}U^\dagger_{x+\nu, x+\mu+\nu}U^\dagger_{x, x+\nu}.
\end{eqnarray}
The two couplings in~\eq{action} are necessary in order to keep physics unchanged under independent variations of $a$ and $\xi$. They are parametrized as:
\begin{eqnarray}
\beta_\sigma = \frac{z}{g^2_\sigma(a, \xi)}\xi^{-1}~~
\text{  and  }~~\beta_\tau = \frac{z}{g^2_\tau(a, \xi)} \xi.
\end{eqnarray}
We will use $z = 2$ for $SU(N)$ and $U(N)$ groups, and $z=1$ for $U(1)$ to ensure the canonical kinetic term in the continuum limit.
The speed of light is defined as $c = g_\sigma/g_\tau$. We will denote the two couplings as $g_\mu$, with $g_\mu = g_\sigma (g_\tau)$ for $\mu$ in the spatial direction (temporal direction). In the weak-coupling limit, the $g_\mu(a,\xi)$ can be expanded in terms of the isotropic value $\beta =z g^{-2}_E(a)$ as:
\begin{eqnarray}
\frac{1}{g^2_\mu(a, \xi)} =  \frac{1}{g^2_E(a)} + c_\mu(\xi) + \mathcal{O}(g^2_E, \xi) 
\label{eq:gmu}
\end{eqnarray}
and $\xi =1$ returns the usual isotropic formulation of a lattice gauge theory with $g_\sigma =g_\tau=g_E$. 
In the weak-coupling regime, the speed of light is given by:
\begin{eqnarray}
c =\frac{g_\sigma(a, \xi)}{g_\tau(a,\xi)}.
\label{eq:cfactor}
\end{eqnarray}

In a more symmetric fashion, the action of \eq{action} can also be rewritten as:
\begin{eqnarray}
S(U) = \frac{z}{g^2_\xi} \sum_x \bigg[\bar{\xi}^{-1}\sum_{i>j}\re\tr P_{ij} +\bar{\xi}\sum_{i} \re\tr P_{0i}\bigg]
\end{eqnarray}
where the bare couplings $g^2_\xi = g_\sigma g_\tau \equiv z/\beta_\xi$ and the bare anisotropy $\bar{\xi} = c \xi$ are introduced; for every $(a, \xi)$ pair there is a corresponding pair of bare couplings $(\beta_\xi, \bar{\xi})$. Following \eq{gmu}, we have:
\begin{equation}
    \frac{1}{g^2_\xi} \approx \frac{1}{g^2_E(a)} + \frac{c_\tau(\xi) + c_\sigma(\xi)}{2}.
\end{equation}
The functions $c_\tau(\xi)$ and $c_\sigma(\xi)$ can be found by calculating the effective action $S^{(\xi)}_{\rm eff}$ of the lattice gauge theory for the two different lattice regularization procedures with $\xi =1$ and $\xi \neq 1$. Requiring that in the continuum limit the effective action is independent of regularization, we have
\begin{eqnarray}
\Delta S_{\rm eff} = S^{(\xi = 1)}_{\rm eff} - S^{(\xi\neq 1)}_{\rm eff} = 0.
\end{eqnarray}
This leads to the determination of $c_\tau(\xi)$ and $c_\sigma(\xi)$. The effective action can be perturbatively calculated using the background field method on the lattice~\cite{Dashen:1980vm}.

We will denote $B_\mu(x)$ as the background field that solves the classical lattice equation of motion. With the fluctuating field $\alpha_\mu$,  the lattice gauge variables can be parametrized as:
\begin{eqnarray}
U_{x, x+\mu} &=& e^{i u g_E a_\mu \alpha_\mu(x)}U^{(0)}_{x, x+\mu}\notag\\
U^{(0)}_{x, x+\mu} &=& e^{i u a_\mu B_\mu(x)}.
\end{eqnarray}
For general dimensions, the couplings and fields may not be dimensionless, thus we rescale the couplings by a factor of $u = a^{D/2-2}$. Note that for one-loop calculations, we can use the isotropic coupling $g_E$ in these exponents instead of $g_\mu$. The covariant derivatives are defined as:
\begin{eqnarray}
\label{eq:derivative}
D_\mu f(x) = \frac{1}{a_\mu}(U_{x, x+\mu}f(x+\mu)U^\dagger_{x, x+\mu}-f(x)),\notag\\
\overline{D}_\mu f(x) = \frac{1}{a_\mu}(U^\dagger_{x-\mu, x}f(x-\mu)U_{x-\mu, x}-f(x)).
\end{eqnarray}
The lattice derivatives $\Delta_\mu f(x)$ and $\overline{\Delta}_\mu f(x)$ follow from~\eq{derivative} by setting $U_{x, x+\mu} = \mathbb{1}$. Taking $U_{x, x+\mu}\rightarrow U^{(0)}_{x, x+\mu}$ defines $D^{(0)}_\mu f(x)$, $\overline{D}^{(0)}_\mu f(x)$. The lattice action can be expanded around $U_{x, x+\mu} = U_{x, x+\mu}^{(0)}$ as:
\begin{equation}
    S(U) = S_0 + S_2 + ...
\end{equation}
where $S_0 = S(U^{(0)})$ and $S_2$ includes terms quadratic in $\alpha_\mu$. To preserve the gauge symmetry of the background field, we work in the background Feynman gauge~\cite{Capitani:2002mp} which requires adding the gauge-fixing term
\begin{eqnarray}
\label{eq:gfix}
S_{\rm gf} = a^{D-1} a_\tau \sum_x\tr\bigg(\sum_\mu \overline{D}^{(0)}_\mu\alpha_\mu(x)\bigg)^2
\end{eqnarray}
and an associated ghost term $S_{\rm gh}(\phi)$ for a ghost field $\phi$ when a non-abelian gauge theory is considered:
\begin{eqnarray}
S_{\rm gh} = 2 a^{D-1} a_\tau \sum_{x,\mu} \tr[(D^{(0)}_\mu\phi(x))^\dagger (D^{(0)}_\mu\phi(x))].
\label{eq:ghost}
\end{eqnarray}
The partition function can be calculated as:
\begin{eqnarray}
\label{eq:effaction}
    Z&\equiv& \int[dU] e^{-S(U)}\notag\\
    &\approx& e^{-S_0}\int [d\alpha][d\phi] e^{-(S_2+ S_{\rm gf}+S_{\rm gh})}\bigg(1+\mathcal{O}(g_E^2)\bigg)\notag\\ 
    &\approx& e^{-S_0}\int[d\phi]e^{-S_{\rm gh}}\int [d\alpha] e^{-S_{\rm free}} e^{-S'_2}\notag\\ 
    &\approx& \int[d\phi] e^{-S_{\rm gh}}\int [d\alpha]e^{-S_{\rm free}}\left(1 - S_0 - S'_2 + \frac{S_2^{'2}}{2} + \hdots\right)\notag\\
    &\propto& e^{-S^{(\xi)}_{\rm eff}} \approx 1 - S^{(\xi)}_{\rm eff} + ...,
\end{eqnarray}
where we have extracted the free action $S_{\rm free}$ for the fluctuating field $\alpha_\mu$ from $S_2 + S_{\rm gf}$ and denote the rest as $S'_2$. On the fourth line, we have Taylor expanded $e^{-S_0-S'_2}$. 

In this article, we consider the $F^2_{\mu\nu}$ term in $S_{\rm eff}(\xi)$ at one loop. This gives the $\mathcal{O}(g_E^0)$ corrections $c_\tau(\xi)$ and $c_\sigma(\xi)$. Matching terms in \eq{effaction} we see $S^{(\xi)}_{\rm eff}$ is related to expectation values computed with respect to $S_{\rm free}$:
\begin{eqnarray}
\label{eq:seffgen}
    S^{(\xi)}_{\rm eff} =&& S_0 + \left<S'_2\right> - \frac{1}{2}\langle S_2^{'2}\rangle \notag\\
    &&+ \left<S_{\rm gh}\right>_\phi,
\end{eqnarray}
Similarly, the contributions from $S_{\rm gh}$ can be calculated as $\left<S_{\rm gh}\right>_\phi$. 
Higher loop corrections carries additional factors of the coupling $g_E^2$ and are negligible at weak coupling. 

\section{\texorpdfstring{$U(1)$}{U(1)} gauge theory}\label{sec:u1}
We apply the background field methods to $U(N)$ and $SU(N)$ to compute the perturbative relations for Euclidean lattices 
at any anisotropy and in any dimension. We will initially consider the simpler $U(1)$ gauge theory, then consider the more involved case of $SU(N)$ gauge theory, followed by the $U(N)$ gauge theory.

For the $U(1)$ gauge theory, $B_\mu$ and $\alpha_\mu$ are the single component electromagnetic fields and we can trivially perform the traces in \eq{gfix} to find
\begin{equation}
    S_{\rm gf} = \frac{1}{2}a^{D-1} a_\tau \sum_{x}\bigg(\sum_\mu \bar{D}^{(0)}_\mu \alpha_\mu(x)\bigg)^2,
\end{equation}
while the $S_{\rm free}$ is found to be
\begin{eqnarray}
S_{\rm free} =\frac{1}{2} a^{D-1} a_\tau \sum_{x, \mu, \nu}(\Delta_\mu \alpha_\nu)(\Delta_\mu \alpha_\nu),
\end{eqnarray}
and $S'_2$ is given by
\begin{equation}
S'_{2} = -\frac{a^{2D-5}a_\tau}{8}\sum_{x,\mu,\nu,a}(a_\mu a_\nu F_{\mu\nu})^2(\Delta_\mu \alpha_\nu -\Delta_\nu\alpha_\mu)^2.
\end{equation}
The non-vanishing contributions to \eq{seffgen} are given by:
\begin{align}
\label{eq:effaction3}
S^{(\xi)}_{\rm eff} = & S_0 + \left<S'_2\right>,
\end{align}
where in $U(1)$ we can neglecting the $\langle S_2^{'2}\rangle$ term as it only contributes to higher orders of $F^2_{\mu\nu}$.  Further, the ghost term is zero in $U(1)$.
$S^{(\xi)}_{\rm eff}$ for $U(1)$ in arbitrary dimensions can then be written as:
\begin{eqnarray}
\label{eq:seff_f}
S^{(\xi)}_{\rm eff} = \frac{1}{4}\int d^D x \bigg(\sum_{i}&&[(F^a_{i0})^2 + (F^a_{0i})^2][g^{-2}_{\tau}- f_{\tau}(\xi)] \notag\\
&&+\sum_{i,k} (F^a_{ik})^2[g^{-2}_{\sigma} - f_{\sigma}(\xi)]\bigg)
\end{eqnarray}
with
\begin{eqnarray}
\label{eq:u1fun}
    f_{\tau}(\xi)&=& \frac{1}{2\xi}\bigg(1-\frac{D-2}{D-1}\xi^{-1}I_1(\xi)\bigg), \notag\\
    f_{\sigma}(\xi) &=& \frac{I_1(\xi)}{D -1}.
\end{eqnarray}
$I_1(\xi)$ and other integrals required for this paper are defined in Appendix~\ref{sec:append}, following~\cite{Karsch:1982ve}. One can show $I_1(1) = \frac{D-1}{D}$ and $f_{\tau}(\xi\rightarrow \infty)=0$. For $\xi=1$, both $f_\mu(1)=1/D$ and thus $g^2_E(\text{one-loop}) = g^2_E[1+ f_{\tau}(1)] = g^2_E[1+ 1/D]$ which agrees with previous $D=4$ calculations~\cite{Cella:1997hw}.
From $f_{\mu}(\xi)$, we obtain
\begin{eqnarray}
\label{eq:u1result}
    &c_\mu(\xi) = f_\mu(\xi) -f_\mu(\xi=1).
\end{eqnarray}
These functions are shown in Fig.~\ref{fig:U1} for 3 and 4 dimensions. In the $\xi\rightarrow \infty$ limit, we show in Appendix~\ref{apx:series} that
\begin{align}
    I_1(\xi\to \infty) \approx \sqrt{\frac{D-1}{2}-\frac{1}{16}}
\end{align}
and that $c_{\tau}(\xi\rightarrow \infty) =-1/D$. Specific numerical values when $D=3,4$ are:
\begin{eqnarray}
    c_{\tau}(\xi\rightarrow \infty) &&=-\frac{1}{3},\,-\frac{1}{4}~~(D=3,~4)
    \notag\\
    c_{\sigma}(\xi\rightarrow\infty) &&= 0.146,\,0.148~~(D=3,~4).
\end{eqnarray}

\begin{figure}
    \centering
    \includegraphics[width=0.92\linewidth]{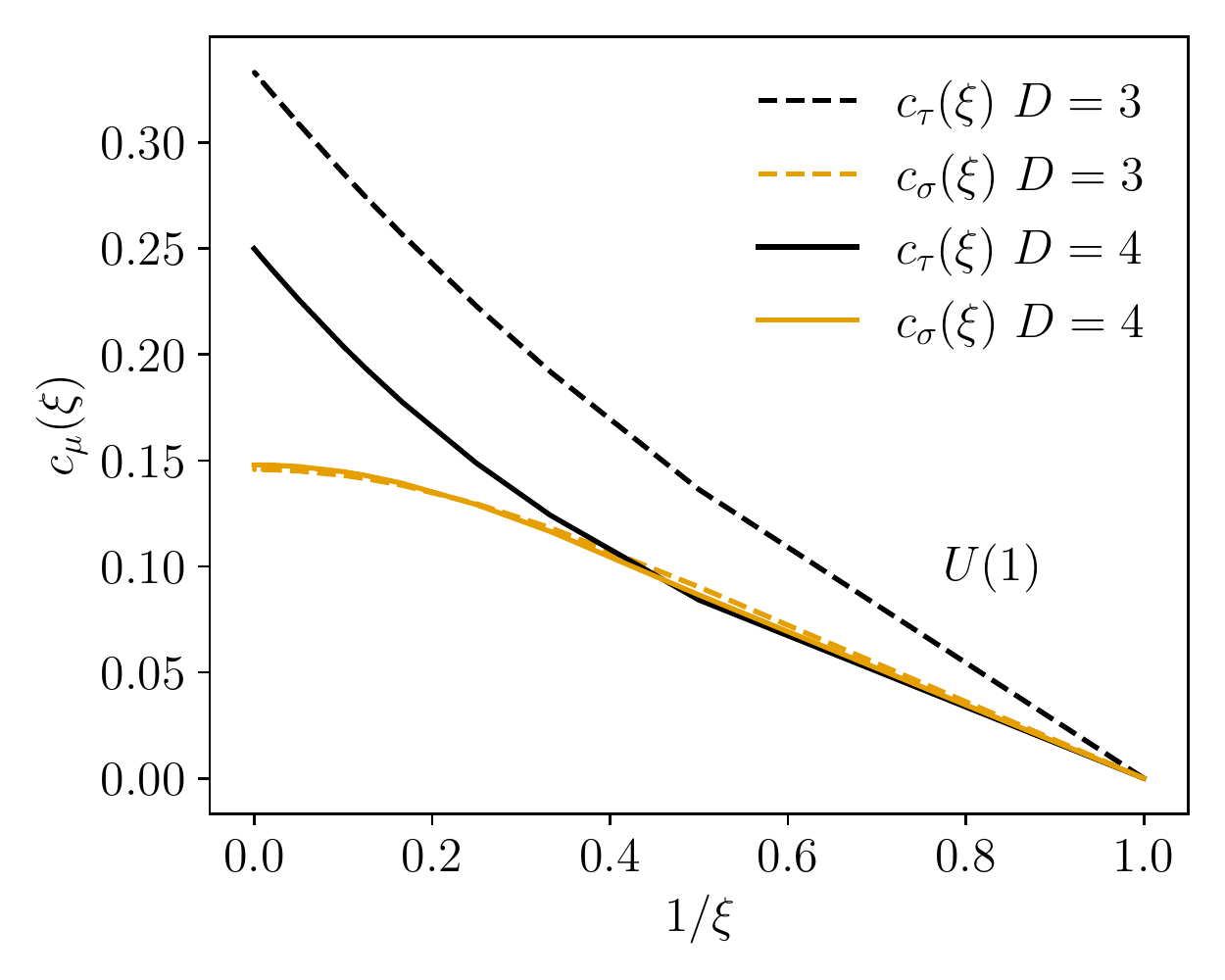}
    \caption{\label{fig:U1} Anisotropic coefficients for $U(1)$ in $D=3,4$.}
\end{figure}

\section{\texorpdfstring{$SU(N)$}{SU(N)} gauge theory}\label{sec:sun}
 We now move to consider the more complicated case of $SU(N)$. $B_{\mu}$ and $\alpha_{\mu}$ can be expanded in terms of the group generators $\lambda^a, a = 1, ..., N^2-1$:
\begin{eqnarray}
 B_\mu = B^a_\mu \lambda^a/2,~~~\alpha_\mu = \alpha^a_\mu \lambda^a/2, 
\end{eqnarray}
with the generators normalized to $\tr \lambda^a\lambda^b = 2\delta_{ab}$. Using
the gauge-fixing term in \eq{gfix}
and the ghost term in \eq{ghost},
we can rewrite $S_2 + S_{\rm gf}$ in terms of a tensor $S_T$, a scalar $S_{\rm sc}$ and two vector interactions $S_A$ and $S_B$:
\begin{align}
S_{T} &= -\frac{a^{2D-5}a_\tau}{8 N}\sum_{x,\mu,\nu,a}(a_\mu a_\nu F^a_{\mu\nu})^2\tr(\Delta_\mu \alpha_\nu -\Delta_\nu\alpha_\mu)^2 \notag\\
S_{\rm sc} &= a^{D-1} a_\tau \sum_{x, \mu, \nu}\tr[(D^{(0)}_\mu \alpha_\nu)(D^{(0)}_\mu \alpha_\nu)]\notag\\
S_{A} &= a^{D-1} a_\tau\sum_{x,\mu\nu}a^{(D-4)/2}\tr(A_{\mu\nu}(x)F_{\mu\nu}(x)) \notag\\
S_{B}  &= \frac{1}{2}a^{D-1}a_\tau\sum_{x,\mu,\nu}a^{(D-4)/2}\tr(B_{\mu\nu}(x)F_{\mu\nu}(x)).
\end{align}
$A_{\mu\nu}(x)$ and $B_{\mu\nu}(x)$ are anti-symmetric and symmetric in the vector indices, respectively, and given by:
\begin{eqnarray}
A_{\mu\nu}(x) = &&-i\{2[\alpha_\nu, \alpha_\mu] + a_\nu[\alpha_\nu, D^{(0)}_\nu\alpha_\mu]\notag\\ 
&&+ a_\mu[D^{(0)}_\mu\alpha_\nu, \alpha_\mu] + \frac{a_\mu a_\nu}{2}[D^{(0)}_\mu\alpha_\nu, D^{(0)}_\nu\alpha_\mu]\}\notag\\
B_{\mu\nu}(x)=&& -i (a_\mu[D^{(0)}_\nu \alpha^\mu, \alpha^\mu] + a_\nu[\alpha^\nu, D^{(0)}_\mu \alpha^\nu]).
\end{eqnarray}
From $S_{\rm sc}$, we extract the free action for the $\alpha_\mu$ field:
\begin{eqnarray}
\label{eq:free}
S_{\rm free} = a^{D-1} a_\tau \sum_{x, \mu, \nu}\tr[(\Delta_\mu\alpha_\nu)(\Delta_\mu\alpha_\nu)]
\end{eqnarray}
and define $S_{\rm sc,I} = S_{\rm sc} - S_{\rm free}$.
The non-vanishing contributions to the effective action are given by:
\begin{align}
\label{eq:effaction2}
S^{(\xi)}_{\rm eff} = & S_0 + \left<S_T\right>-\frac{1}{2}\left<S^2_A\right> -\frac{1}{2}\left<S^2_B\right>\notag\\ &+\frac{D-2}{D}\left[\left<S_{{\rm sc},I}\right>-\frac{1}{2}\left<S^2_{{\rm sc},I}\right>\right]
\end{align}
with other terms vanishing. Notice that unlike $U(1)$, $\langle S_2^2\rangle$ terms contribute at leading order. The factor $\frac{D-2}{D}$ comes from the fact that the ghost field contribution cancels $2$ out of $D$ degrees of freedom of $\alpha_\mu$. Where, again the expectation values are calculated with respect to $S_{\rm free}$.  The final one-loop corrected action is given by:
\begin{eqnarray}
\label{eq:seff_f2}
S^{(\xi)}_{\rm eff} = \frac{1}{4}\int d^D x \bigg(&&\sum_{i}[(F^a_{i0})^2 + (F^a_{0i})^2][g^{-2}_{\tau}- f_{\tau}(\xi)] \notag\\
&&+\sum_{i,k} (F^a_{ik})^2[g^{-2}_{\sigma} - f_{\sigma}(\xi)].
\end{eqnarray}
$SU(N)$, $f_\mu(\xi)$ are defined as,
\begin{eqnarray}
\label{eq:sufun}
&&f_\tau(\xi) =4 N\bigg[ \frac{N^2-1}{16N^2}\{\frac{\xi^{-2} I_1(\xi)}{(D-1)} + \xi^{-1}I_5(\xi)\} +\frac{1}{64}I_{2b}(\xi) \notag\\&& + \frac{D-14}{384(D-1)}I_{2a}(\xi) + \frac{1}{256}I_4(\xi)\xi^{-2} + \frac{D-8}{192}\xi^{-2} I_6(\xi)\notag\\&& +\frac{2-D}{384}\xi^{-2} I_7(\xi) +\frac{26-D}{24}{\rm DIV}(\xi)\bigg],\notag\\
\notag\\
&&f_\sigma(\xi) = 4 N\bigg[\frac{N^2-1}{8N^2}\frac{I_1(\xi)}{(D-1)} + 
\frac{D-14}{192(D-1)}I_{2a}(\xi)\notag\\&&+\frac{8-D}{192} I_3(\xi) +\frac{1}{128}I_4(\xi)+\frac{26-D}{24}{\rm DIV}(\xi)\bigg].
\end{eqnarray}
The DIV part defined as:
\begin{eqnarray}
    {\rm DIV}(\xi) &&= \frac{2^{D-4}}{(2\pi)^D}\int^{\pi/2}_{-\pi/2}d^{D-1}x \int^{(\pi/2)\xi}_{(-\pi/2)\xi}dx_0\notag\\
    && \bigg(\sum^{D-1}_{i=1}\sin^2 x_i + \xi^2\sin^2(x_0/\xi)\bigg)^{-2},
\end{eqnarray}
is divergent. This divergence comes from $S_A$ and $S_{\rm sc, I}$ terms which do have corresponding continuum limit and contain logarithmic divergence as $a\rightarrow 0$. With the definition for $c_\mu(\xi)$ in \eq{u1result}, the divergence part in ${\rm DIV}(\xi)$ are subtracted out.
\begin{figure}
    \centering
    \includegraphics[width=0.92\linewidth]{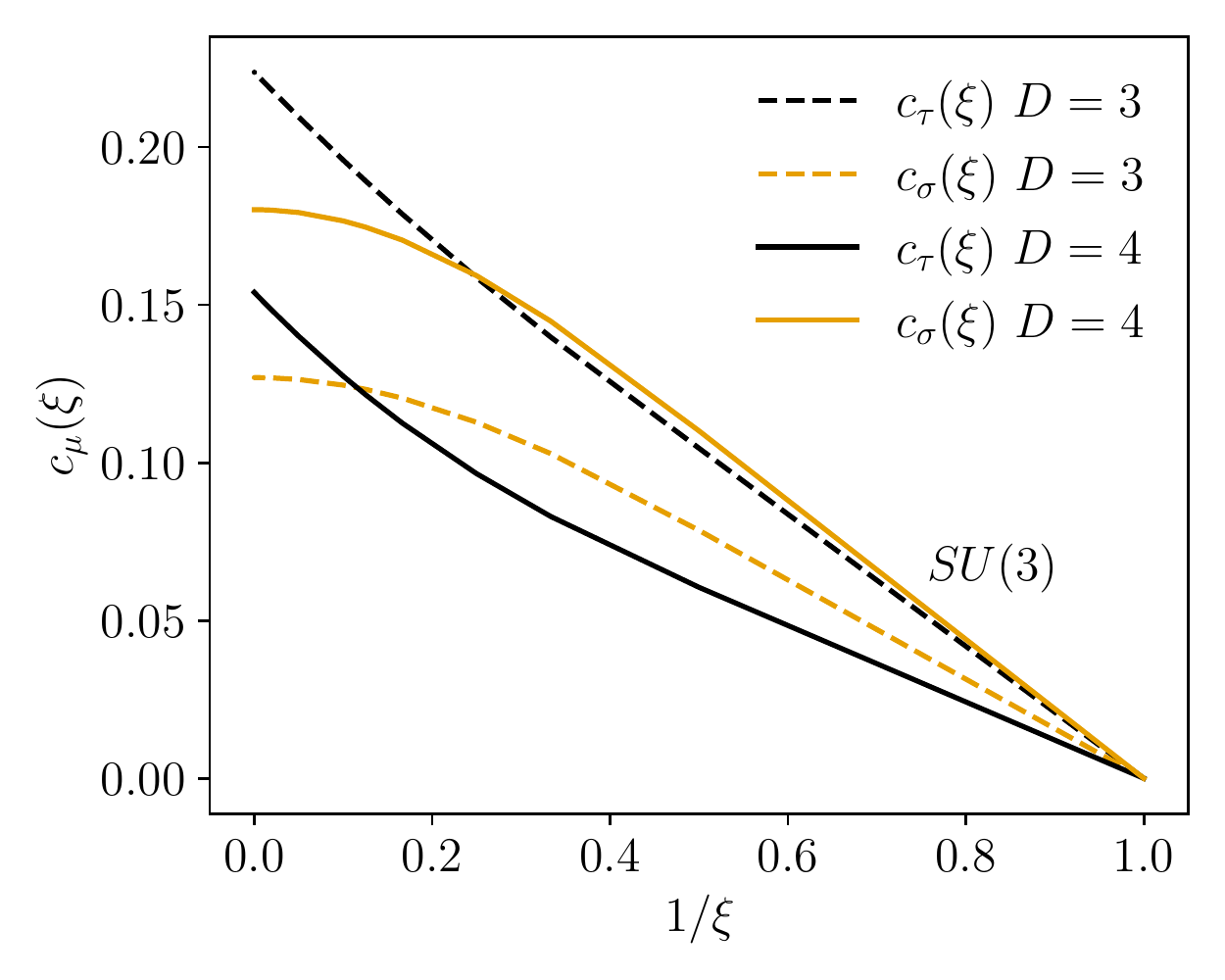}
    \caption{\label{fig:SU3} Anisotropic coefficients for $SU(3)$ in $D=3,4$.}
\end{figure}
Our calculation gives the same results as~\cite{Karsch:1982ve} for $D=4$ and as~\cite{Hasenfratz:1981tw, Hamer:1996ub} in the $\xi\rightarrow \infty$ limit. The values of $c_\mu(\xi)$ for $SU(3)$ gauge theory are shown in~\fig{SU3} at different dimensions.

\section{\label{sec:un} \texorpdfstring{$U(N)$}{U(N)} gauge theory}
The Lie algebra for $U(N)$ group can be constructed by introducing the additional generator $\lambda^0 = \sqrt{\frac{2}{N}}\mathbb{I}_{N\times N}$ to the $SU(N)$ group. Corresponding to any index $a$ for $SU(N)$ group we introduce the index $A = (0, a)$, so that $A$ runs from 0 to $N^2-1$. With this construction, we still have $\tr[\lambda^A \lambda^B] = 2\delta_{AB}$; special care has to be taken for the anti-symmetric structure constant as $f_{0BC} = 0$. The final one-loop corrected action is given by:
\begin{eqnarray}
\label{eq:seff_f3}
S^{(\xi)}_{\rm eff} = \frac{1}{4}\int d^D x \bigg(&&\sum_{i}[(F^a_{i0})^2 + (F^a_{0i})^2][g^{-2}_{\tau}- f_{\tau}(\xi)] \notag\\
&&+\sum_{i,k} (F^a_{ik})^2[g^{-2}_{\sigma} - f_{\sigma}(\xi)]\notag\\&&+\sum_{i}[(F^0_{i0})^2 + (F^0_{0i})^2][g^{-2}_{\tau}- f_{0,\tau}(\xi)] \notag\\
&&+\sum_{i,k} (F^0_{ik})^2[g^{-2}_{\sigma} - f_{0,\sigma}(\xi)]\bigg),
\end{eqnarray}
with $f_\tau(\xi)$ and $f_\sigma(\xi)$ given by \eq{sufun} but replacing $N^2-1$ by $N^2$ which changes the factor $\frac{N^2-1}{N^2}$ to 1,
 and $f_{0, \tau}(\xi)$ and $f_{0,\sigma}(\xi)$ corresponding to Eq.~(\ref{eq:u1fun}) multiplied by a factor of $N/2$.

\section{\label{sec:discretegroup}Comparing to Numerical Results}
Our values of $c_\sigma$ and $c_\tau$ computed in Sec.~\ref{sec:u1} and Sec.~\ref{sec:sun} can be used to calculate the renormalized anisotropy, $\xi$ using the relation $\bar{\xi} = c\xi$, with $c$ given in \eq{cfactor} and expressions for $g_\mu$ in \eq{gmu}.
In this section, we compare our one-loop calculations of $\xi$ as well as the bare anisotropy with nonperturbative results obtained from two sets of Monte Carlo results.  The first are existing 2+1d $U(1)$ and $SU(2)$ results produced in Refs. \cite{Loan:2003wy,Teper:1998te}.  The second are new ensembles produced by us for the discrete cyclic groups $\mathbb{Z}_{10}$ and $\mathbb{Z}_{100}$, and the binary icosahedral ($\bi$).  These discrete groups are of interest because they are subgroups of $U(1)$ and $SU(2)$, respectively, and have been proposed as approximations for use on quantum computers. Thus, it is interesting to investigate how well perturbative lattice field theory for the continuous group can approximate $\xi$ for the discrete subgroups. In both previous works, smearing was used to reduce the need for higher statistics. This has the consequence of changing the lattice spacings by a unknown, potentially large amount and can introduce some discrepancy between the perturbative and the nonperturbative results.  For this reason, in our simulations we avoided using smearing at the cost of a larger number of lattice configurations.

\begin{table}
\caption{Ensemble parameters for the lattice simulations: Group $G$, coupling $\beta$, bare anisotropy $\bar{\xi}$, Lattice dimensions $N_s^D\times N_t$, decorrelation length $n_{\text{decor}}$ and number of configurations $n_{\text{meas}}$.}
\label{tab:ensembleparams}
\centering
\begin{tabular}{c r c c c c}
    \toprule
    $G$&$\beta$&$\bar{\xi}$&$N_s^D\times N_t$&$n_{\rm decor}$&$n_{\rm meas}$\\
    \hline
    $\mathbb{Z}_{10}$, $\mathbb{Z}_{100}$&1.35&2.25&$16^2\times48$&10&$8\times10^6$\\
    $\mathbb{Z}_{10}$, $\mathbb{Z}_{100}$&1.35&2.25&$24^2\times72$&10&$4\times10^6$\\
    $\mathbb{Z}_{10}$, $\mathbb{Z}_{100}$&1.55&2.5&$16^2\times48$&10&$5\times10^6$\\
    $\mathbb{Z}_{10}$, $\mathbb{Z}_{100}$&1.55&2.5&$24^2\times72$&10&$1\times10^6$\\
    $\mathbb{Z}_{10}$, $\mathbb{Z}_{100}$&1.7&3.0&$16^2\times48$&10&$5\times10^6$\\
    $\mathbb{Z}_{10}$, $\mathbb{Z}_{100}$&1.7&3.0&$20^2\times60$&10&$5\times10^6$\\
    $\mathbb{Z}_{10}$, $\mathbb{Z}_{100}$&1.7&3.0&$24^2\times72$&10&$1\times10^6$\\
    $\mathbb{Z}_{10}$, $\mathbb{Z}_{100}$&2.0&3.0&$16^2\times48$&10&$5\times10^6$\\
    $\mathbb{Z}_{10}$, $\mathbb{Z}_{100}$&2.0&3.0&$20^2\times60$&10&$5\times10^6$\\
    $\mathbb{Z}_{10}$, $\mathbb{Z}_{100}$&2.0&3.0&$24^2\times72$&10&$1\times10^6$\\
    $\mathbb{BI}$ & 2.0 & 2.0 & $36^2\times72$ &10 & $5000$ \\
    $\mathbb{BI}$ & 3.0 & 1.33 & $36^2\times72$ & 10& $5000$\\
    $\mathbb{BI}$ & 3.0 & 1.33 & $36^3\times72$ & 10& $250$ \\
    \hline
\end{tabular}
\end{table}

\begin{figure}
    \centering
    \includegraphics[width=\linewidth]{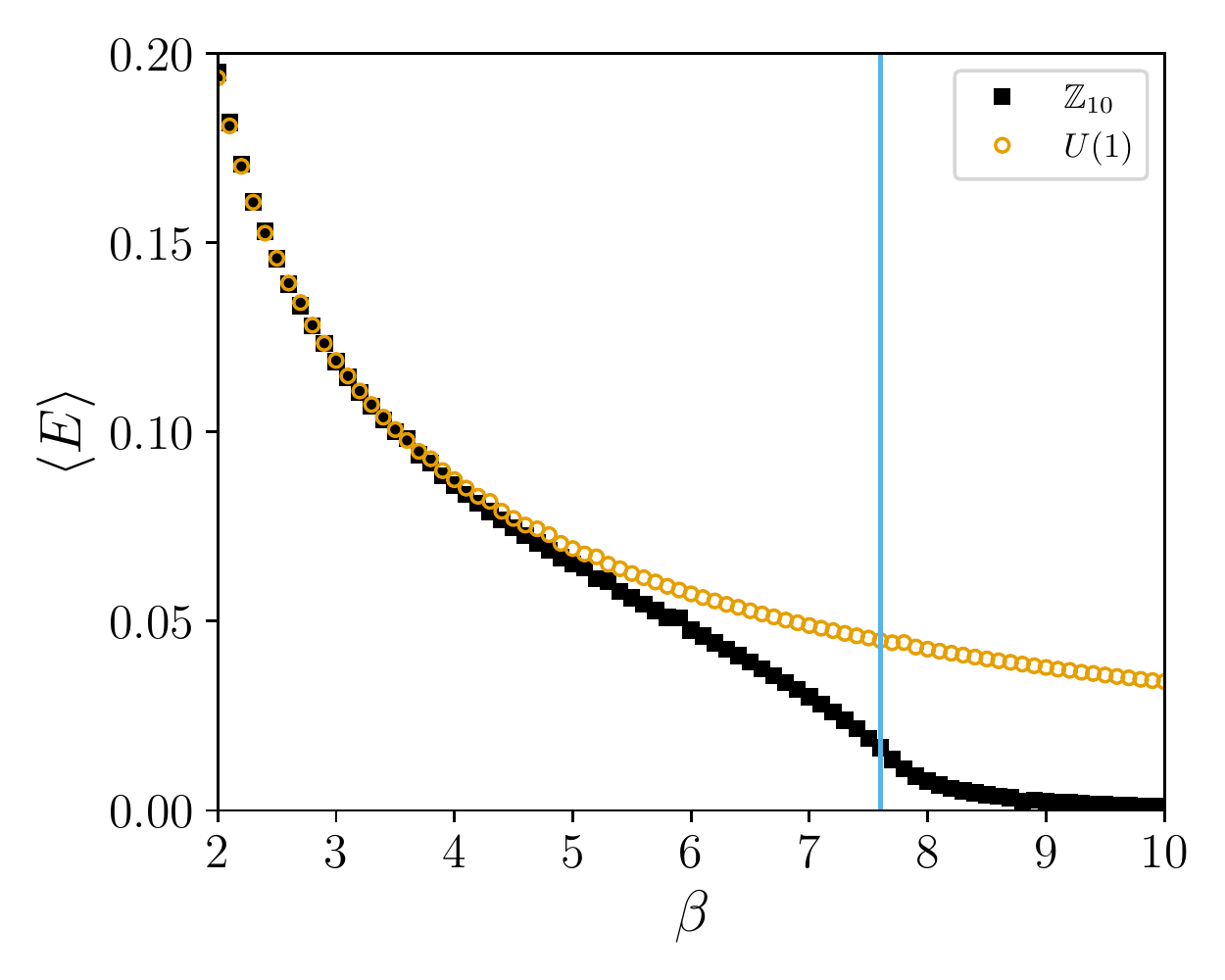}
    \caption{Average plaquette energy $\langle E\rangle$ as a function of $\beta$ for $Z_{10}$ and $U(1)$, with  $\beta_{f,10}=7.6$ indicated by the vertical line.}
    \label{fig:z10}
    \end{figure}

\begin{figure}
    \centering
    \includegraphics[width=\linewidth]{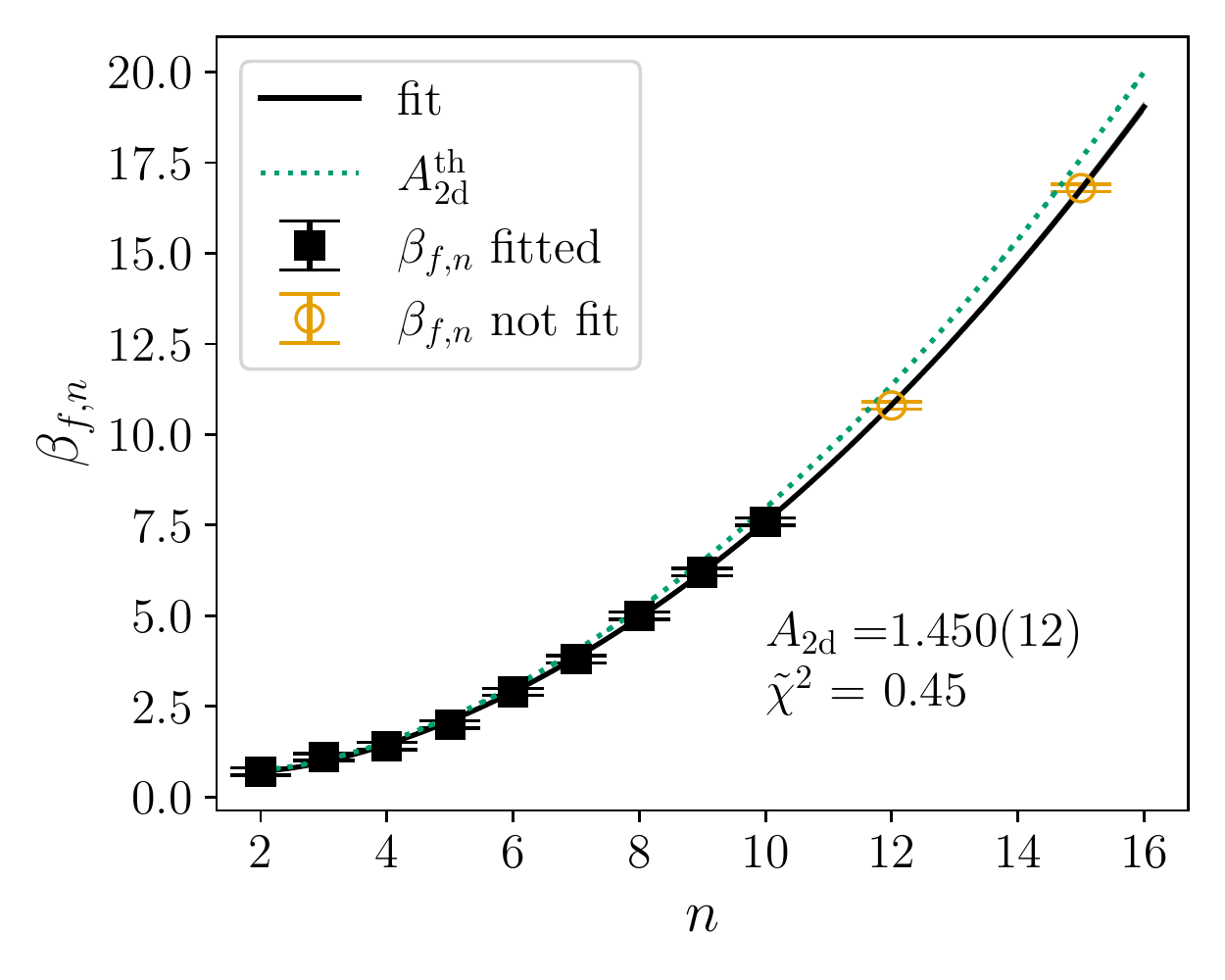}
    \caption{$\beta_{f,n}$ versus $n$. $\mathbb{Z}_n$ for $n\leq10$ (black square) were used to perform the fix, while $n>10$ to test the extrapolation.}
    \label{fig:detA}
    \end{figure}

The discrete group ensembles were generated by sampling from the Wilson action using a multi-hit Metropolis update algorithm, which has been found to be as efficient as a heat-bath in terms of autocorrelation length but significantly cheaper to implement for discrete groups~\cite{Alexandru:2019nsa}. The various ensemble parameters are found in Tab.~\ref{tab:ensembleparams}. Using discrete groups, we must worry about crossing into the frozen phase where all the links take the value of group identity $\mathbb{1}$ at a critical coupling $\beta_f$ for isotropic lattice. For $\mathbb{Z}_n$ groups, it is known that~\cite{Petcher:1980cq}:
\begin{equation}
\label{eq:bfn}
    \beta_{f,n}\approx\frac{A}{1-\cos\left(\frac{2\pi}{n}\right)},
\end{equation}
For $3+1d$ in \cite{Petcher:1980cq}, the theoretical value of $A$ was obtained to be $A^{\rm th}_{\rm 3d}\approx\log(1+\sqrt{2})$, while numerical simulations gave $A_{\rm 3d}=0.78$\cite{Creutz:1979kf,Creutz:1979zg}.
For the case of $2+1d$, using the value $\beta_{f, 2} = 0.761412$ obtained from Monte Carlo simulations in \cite{Hasenbusch:1992zz,Caselle:1994df,Agostini:1996xy}, the theoretical value of $A$ is calculated following \eq{bfn} to be $A^{\rm th}_{\rm 2d}=1.52282$. As a comparison, we compute $A_{\rm 2d}$ with the following procedure. For certain $n$, we measure the average plaquette energy $\left<E\right>$ as a function of $\beta$. $\beta_{f,n}$ is determined as the $\beta$ value that maximizes  the specific heat $\left|\frac{\partial\left<E\right>}{\partial\beta}\right|$. We compute $\beta_{f, n}$ for $n=2, 3..., 10$ on $10^3$ lattices. As an example, we show the measured value of $\left<E\right>$ in \fig{z10} for $n=10$ at different $\beta$ values. Fitting the $\beta_{f,n}$ values to Eq.~(\ref{eq:bfn}), we obtain $A_{\rm 2d}=1.450(12)$. Two additional $\beta_{f, n}$ for $n=12, 15$ are computed and they agree well with the fit. These results are plotted in \fig{detA}.
Comparing to the $A^{\rm th}_{\rm 2d}$, we expect that corrections to the theoretical value are needed.

In the case of anisotropic lattices considered here, one should expect the effects of the freezing-out to occur when $\beta_{\xi}\bar{\xi} =\beta_{\xi, f}\bar{\xi}   \approx\beta_f$. However, as we observe, for isotropic lattice, $\mathbb{Z}_{10}$ deviates from $U(1)$ around $\beta\approx5$ which is much smaller than $\beta_{f,10}=7.6$ (See Fig.~\ref{fig:z10}). Hence, we expect 
that $\beta_{\xi}\bar{\xi}\ll \beta_f$ is necesssary to ensure discrete subgroups being a reasonable approximation in $2+1d$. 
In $3+1d$, deviations occur at $\beta$ values relatively closer to $\beta_f$ compared to $2+1d$, as observed in~\cite{Alexandru:2019nsa,Alam:2021uuq}.
To compare with existing nonperturbative results for $U(1)$ in $2+1d$ studied by Loan et al \cite{Loan:2003wy}, we generate ensembles for$\mathbb{Z}_{10}$ and $\mathbb{Z}_{100}$ groups at the same set of $(\beta_{\xi},\bar{\xi})$ used by them (see \tab{anisotropyu1}). The two largest pairs investigated in \cite{Loan:2003wy} are $\beta_{\xi}\bar{\xi}=1.7\times3.0=5.1$ and $\beta_{\xi}\bar{\xi}=2.0\times3.0=6.0$, not much smaller than $\beta_{f, 10}$, and therefore we expect to observe breakdown of the agreement between $\mathbb{Z}_{10}$ and the $U(1)$ results. In contrast, $\beta_{f,100}>700$ so $\mathbb{Z}_{100}$ results should be indistinguishable from a equivalent $U(1)$ computation.

The results of 2+1d $SU(2)$ from \cite{Teper:1998te} consider  $\beta_{\xi}\bar{\xi}$ (see \tab{gateperlink}) above or too close to $\beta_{f,\mathbb{BI}}=9.65(1)$ that we calculated by similar procedures described above. Hence in the following we will not make direct comparisons between discrete groups and continuous groups, but instead compare the viability of the one-loop calculations with the $SU(2)$ continuous group. Then we computed $\mathbb{BI}$ configurations at different values where $\beta_{\xi}\bar{\xi}=4$ and compare those results with our one-loop calculations.
Additonally, we performed one simulation of 3+1d $\mathbb{BI}$ and also compare with our one-loop calculations .

Different methods are available for determining $\xi$ from lattice results. Loan et al.~\cite{Loan:2003wy} utilized the ratio of subtracted static potentials, where a subtraction point must be picked. Teper et al. \cite{Teper:1998te} uses two methods:
the first compares correlators in the spatial and temporal direction which can also be used to determine $\xi$ in real-time simulations \cite{Carena:2021ltu}, the second computes the dispersion relation with low-lying momentum states and tunes $\xi$ to obtain $E(p)\approx \sqrt{m_P^2+p^2}$. These two results are then averaged to obtain a final estimate of $\xi$.

\begin{figure}
    \centering
    \includegraphics[width=\linewidth]{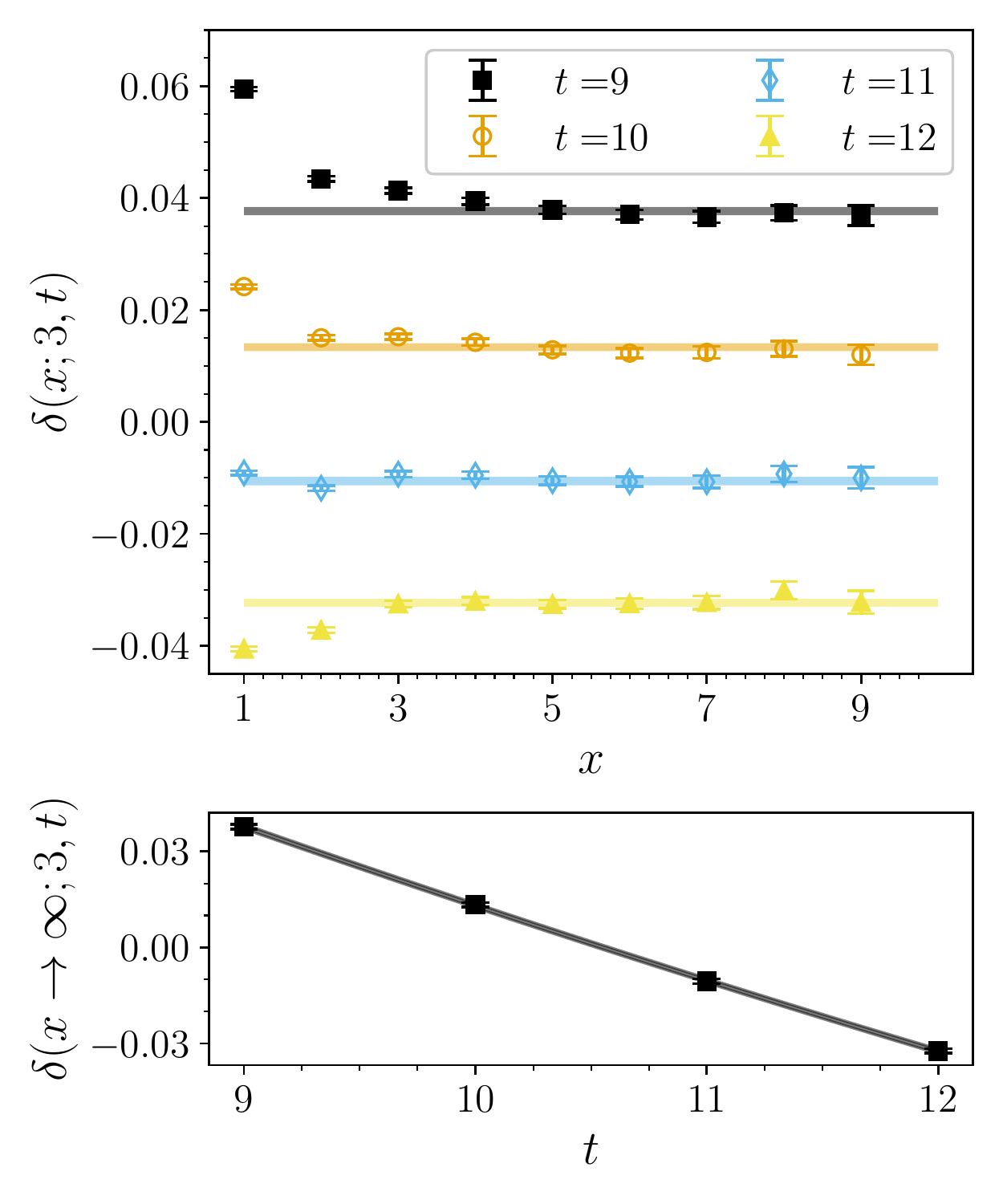}
    \caption{Example calculations of $\delta(x; y, t)$ for $y=3$ as a function of $x$ (top),  $\delta(x\to\infty,3,t)$ for various values of $t$ (bottom) fitted to determine $\delta(x\to\infty, y; t)=0$, for $\mathbb{Z}_{100}$ using $\beta_\xi=1.7$, and $\bar{\xi}=3.0$ on a lattice of size $N_s^D\times N_t=20^2\times60$.}
    \label{fig:interpolation}
    \end{figure}

\begin{figure}
    \centering
    \includegraphics[width=\linewidth]{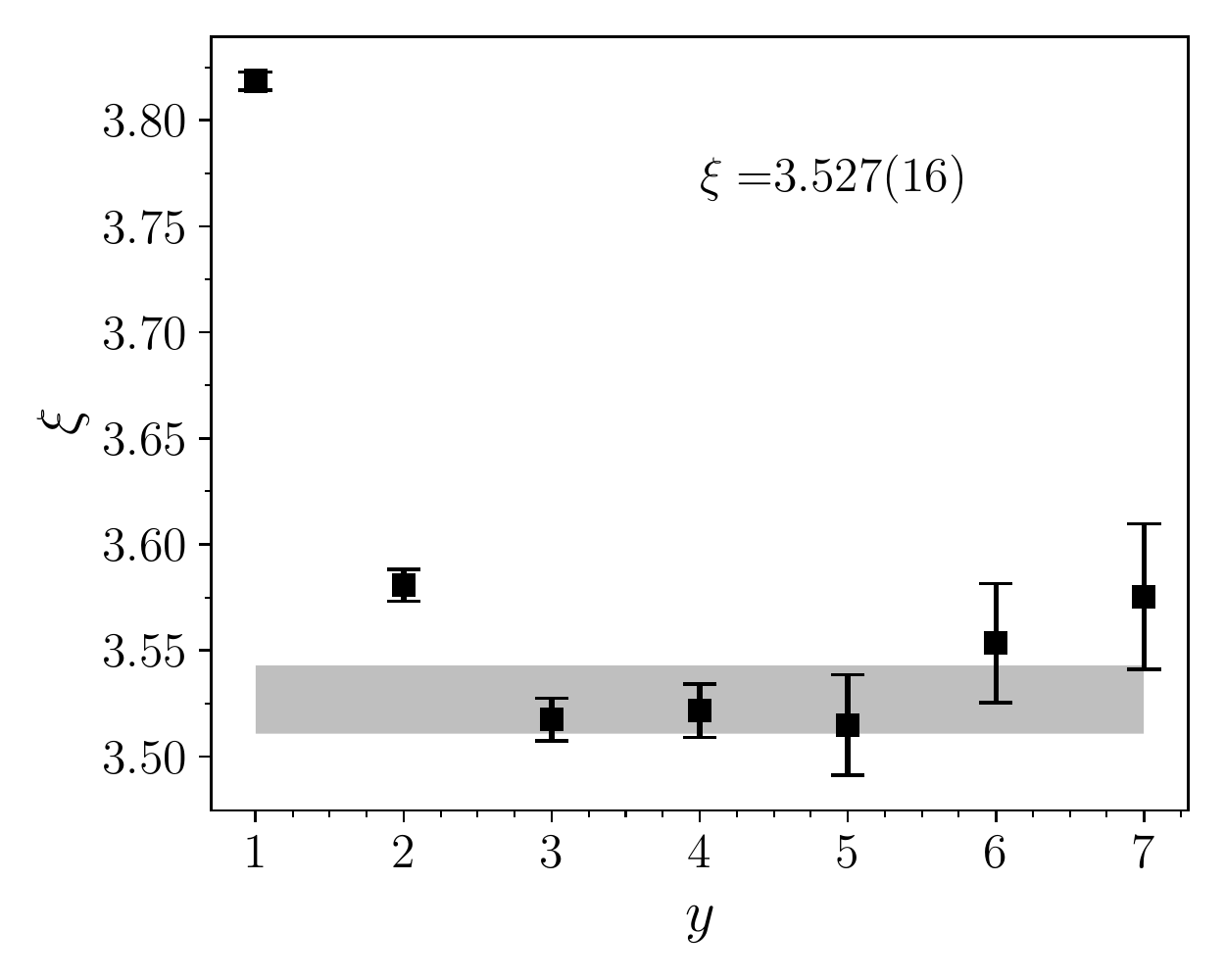}
    \caption{Measured $\xi$ as a function of y. The band corresponds to the 1$\sigma$ error band for best fit to the plateau region. The ensemble parameters are the same as for Fig. \ref{fig:interpolation}.}
    \label{fig:anisotropycalculation}
\end{figure}

We determined $\xi$ for the discrete groups via the `ratio-of-ratios' method~\cite{Klassen:1998ua}. This method involves computing the ratios of Wilson loops:
\begin{equation}
    R_{ss}(x,y)=\frac{W_{ss}(x,y)}{W_{ss}(x+1,y)},\, R_{st}(x,t)=\frac{W_{st}(x,t)}{W_{st}(x+1,t)},
\end{equation}
where $x,y,t$ are the integer lattice separations and the subscripts indicate the orientation of the Wilson loops, either spatial-spatial, or spatial-temporal. In the large $x$ limit where the excited state contamination is suppressed, we have $W_{ss}(x,y)\propto e^{-a x V_s(y a)}$ and $W_{st}(x,t)\propto e^{-a x V_s(t a_\tau)}$ with $V_s$ being the static quark-antiquark potential. This lead to:
\begin{eqnarray}
    R_{ss}(x,y)|_{x\to\infty}=e^{-a V_s(y a)},\\
    R_{st}(x,t)|_{x\to\infty}=e^{-a V_s(t a_\tau)}.
\end{eqnarray}
We define a variable
\begin{equation}
\delta(x,y,t)=\frac{R_{ss}(x,y)}{R_{st}(x,t)}-1,
\end{equation}
such that $\delta(x,y,t)=0$ will be satisfied in the large $x$ limit when $y a = t a_\tau$. We determine $\xi(y) = t/y$. \fig{interpolation}(top) shows the plateau behavior of $\delta(x, y, t)$ when we approach the large $x$ limit.
Typically, the zero crossing does not occur for integer $y,t$ and thus interpolation between values is required. An example of this calculation is shown in Fig.~\ref{fig:interpolation}(bottom) for $\mathbb{Z}_{100}$ using $y=3$, $\beta_{\xi}=1.7$, and $\bar{\xi}=3.0$ on a lattice of size $N_s^D\times N_t=20^2\times60$. The final step is to take the $\xi(y)$ value in the large $y$ limit as our renormalized $\xi$, to again remove excited states contamination (See Fig.~\ref{fig:anisotropycalculation}). The increasing uncertainty at larger $y$ is due to exponential decay of the Wilson loop $W_{ss}(x, y)$ leading to a signal-to-noise problem.

\begin{figure}
    \centering
    \includegraphics[width=\linewidth]{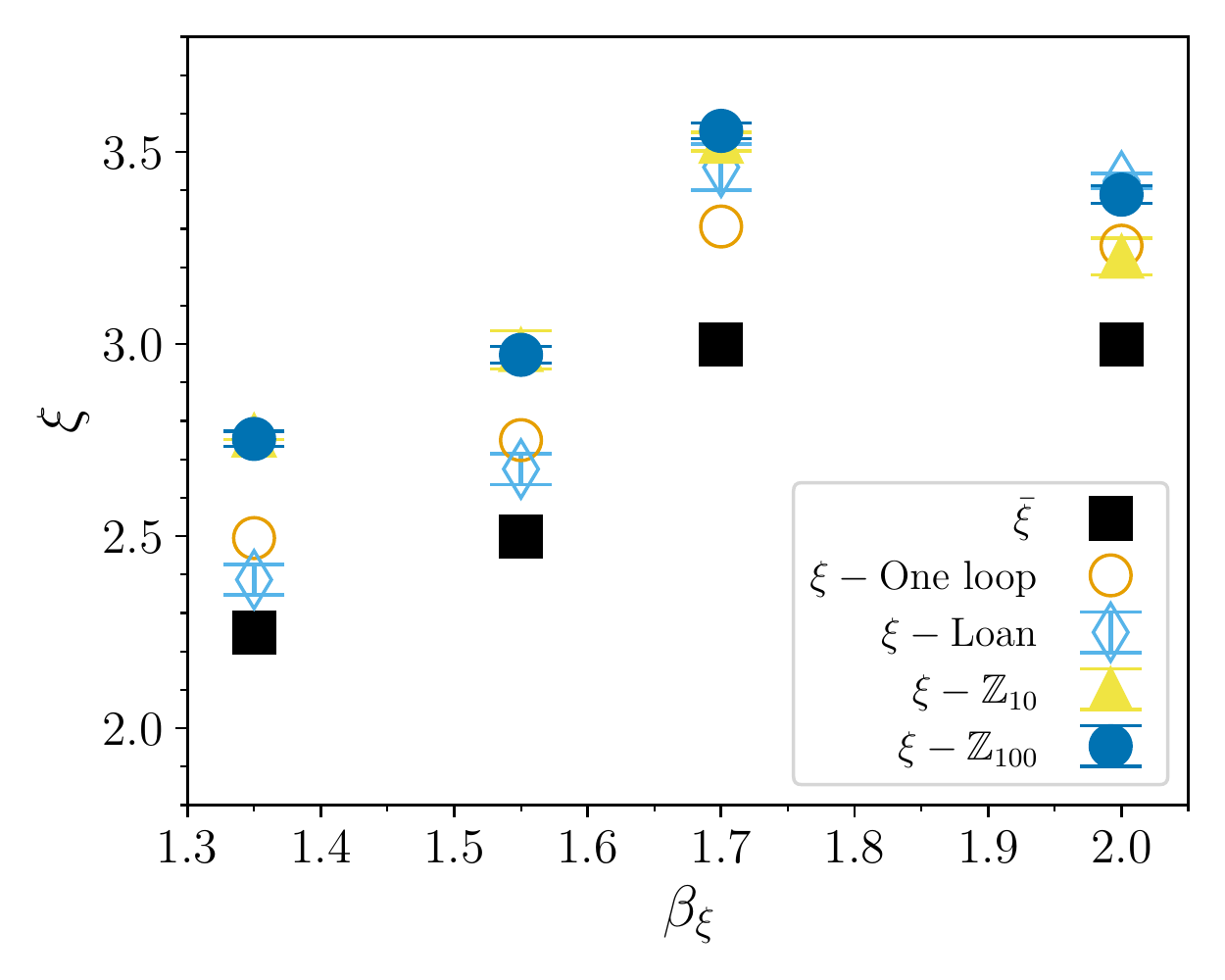}
    \caption{\label{fig:u1matching} Comparison of one-loop $\xi$ to $\bar{\xi}$, the nonperturbative $\xi$ value from Loan et al ($r_0 =\sqrt{2}$)~\cite{Loan:2003wy} for $2+1d$ $U(1)$ theory, and for $\mathbb{Z}_{n}$ discrete group.}
\end{figure}

In \fig{u1matching}, we show the comparisons between the $\xi$ values from our one-loop calculation to the results from non-perturbative Monte Carlo simulation from Loan et al \cite{Loan:2003wy} for $U(1)$ gauge theory in $2+1d$, alongside with our results for $\mathbb{Z}_{10}$ and $\mathbb{Z}_{100}$ also in $2+1d$. It is encouraging to see that including one-loop effects shifts $\xi$ into better agreement with the nonperturbative results compared to $\bar{\xi}$. As a metric for comparison, we use the relative errors:
\begin{eqnarray}
    \mathcal{F}_{g}&=&\left|1-\frac{\xi_{\rm 1-loop}}{\xi_{g}}\right|, \notag\\\mathcal{F}^{\bar{\xi}}_{g}&=&\left|1-\frac{\bar{\xi}}{\xi_{g}}\right|
\end{eqnarray}
where $g$ is the nonperturbative data to which we are comparing our one loop results. For the smeared results for $U(1)$, we find $\mathcal{F}_{U(1)}\leq 4.71(8)\%$ compared to $ \mathcal{F}^{\bar{\xi}}_{U(1)}\leq 13.3(15)\%$.  

The situation with $\mathbb{Z}_n$ is more involved.  For $\beta_{\xi}<1.65$, we find that $\xi_{\mathbb{Z}_{10}}=\xi_{\mathbb{Z}_{100}}$ but are systematically higher than the $U(1)$ results.  At present, we do not understand why at lower $\beta_{\xi}$ greater disagreement is found between the discrete and continuous groups, since lower values of $\beta_{\xi}$ are further from the freezing-out regime and they should be in better agreement. 
We investigated whether finite volume effects could be playing a role, but for all values of $\{\beta_{\xi},\bar{\xi}\}$, we observed no volume-dependence, as see in \tab{anisotropyu1} and \tab{gateperlink}. Two possible sources of the discrepancy could be the use of smearing in \cite{Loan:2003wy,Teper:1998te} or the different methods of measuring $\xi$. Future work should be undertaken where the discrete and continuous groups are analyzed under the same circumstances. As $\beta_{\xi}$ increases, $\xi_{\mathbb{Z}_{100}}$ approaches $\xi_{U(1)}$, with $\mathbb{Z}_{10}$ having noticeable and growing disagreement. 
Across $\beta$, we found $\mathcal{F}_{\mathbb{Z}_{10}}\approx\mathcal{F}_{\mathbb{Z}_{100}}\leq9.5(3)\%$
compared to $\mathcal{F}^{\bar\xi}_{\mathbb{Z}_{10}, \mathbb{Z}_{100}}\leq 18.5(3)\%$.
Higher order loop corrections, of $\mathcal{O}(g^4_E)$ to the $\xi$ could be important for the $\beta_{\xi}$ regions considered and effects of monopoles may also be relevant~\cite{Cella:1997hw}. 

\begin{table}[ht]
    \caption{Renormalized Anisotropies of $U(1)$ from 1-loop calculation, lattice simulations of $\mathbb{Z}_{10}$ and $\mathbb{Z}_{100}$, and $U(1)$~\cite{Loan:2003wy}.}
    \label{tab:anisotropyu1}
    \begin{tabular}{cccc|cccc}
    \hline\hline
    $\beta_{\xi}$& $N_s$ & $N_t$ & $\bar{\xi}$ &$\xi_{\rm 1-loop}$ & \multicolumn{3}{c}{$\xi$}\\\hline
    &&&&&$\mathbb{Z}_{10}$&$\mathbb{Z}_{100}$&$U(1)$~\cite{Loan:2003wy}\\
    \hline
    1.35&16&48&2.25&2.493&2.738(20)&2.732(30)&2.39(4)\\
    1.35&24&72&2.25&2.493& 2.762(10) & 2.753(20) & $\cdots$\\
    1.55&16&48&2.50&2.750&2.939(29)&2.972(40)&2.72(9)\\
    1.55&24&72&2.50&2.750& 2.984(50) &  2.972(22)&$\cdots$\\
    1.70&16&48&3.00&3.302&3.513(11)&3.572(10)&3.46(6)\\
    1.70&20&60& 3.00& 3.302 & 3.512(20)& 3.527(16)&$\cdots$ \\
    1.70&24&72&3.00&3.302& 3.527(24) & 3.555(20)&$\cdots$\\
    2.00&16&48&3.00&3.253&3.259(17)&3.421(15)&3.42(3)\\
    2.00&20&60&3.00&3.253&3.252(38)&3.379(20)&$\cdots$\\
    2.00&24&72&3.00&3.253&3.228(48)&3.389(23)&$\cdots$\\
    \hline
    \end{tabular}
\end{table}

\begin{figure}
    \centering
    \includegraphics[width=\linewidth]{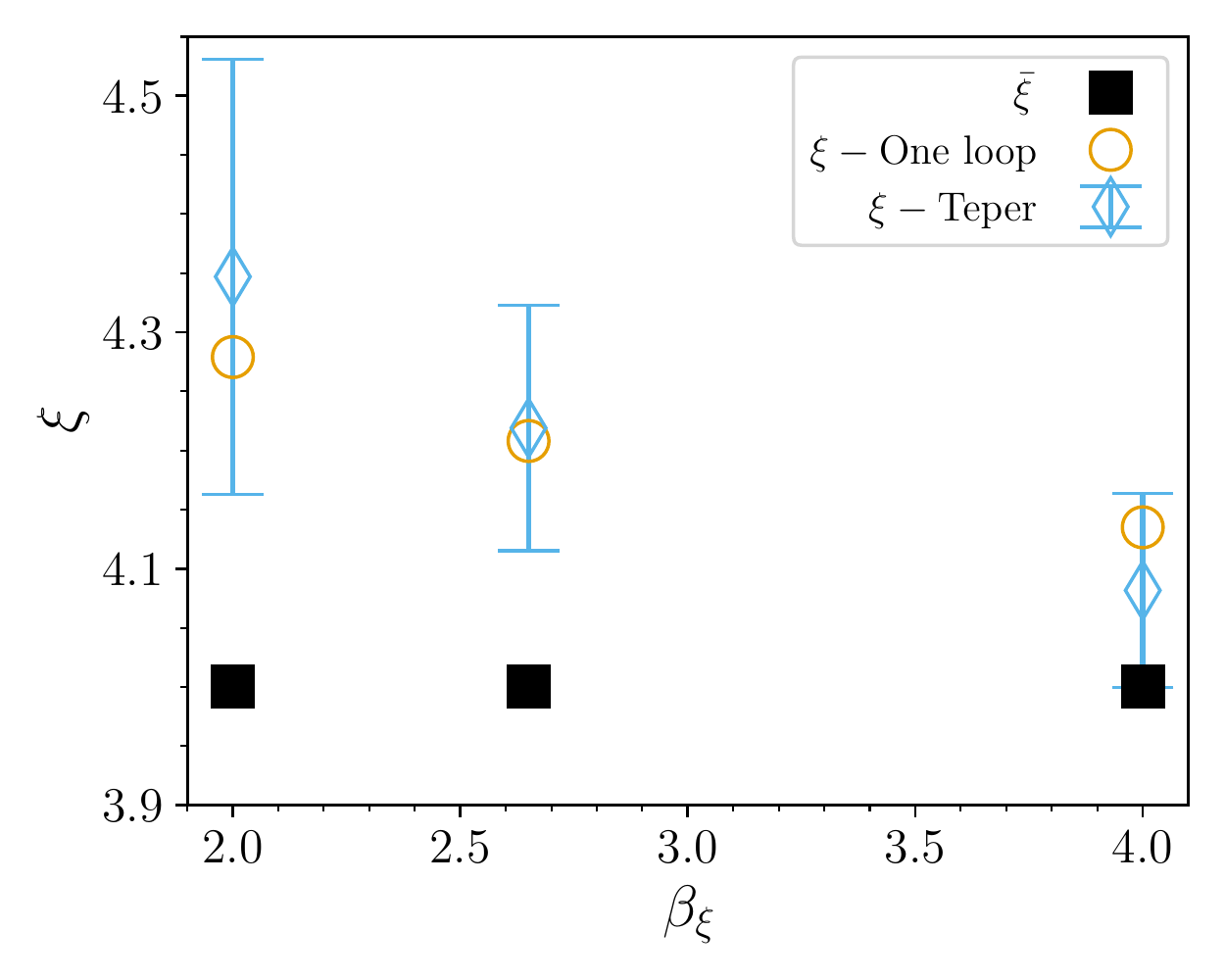}
    \caption{\label{fig:su2matching} Comparison of one-loop $\xi$ to nonperturbative value of~\cite{Teper:1998te} for $2+1d$ $SU(2)$, with $\bar{\xi} = 4$.}
\end{figure}

We can also compare our $\xi_{\rm 1-loop}$ for $SU(2)$ in $2+1d$ to the results from non-perturbative Monte Carlo simulations \cite{Teper:1998te}, which are shown in \fig{su2matching} and \tab{gateperlink}, and to the $\xi$ values we computed for the $\mathbb{BI}$ group (\tab{gateperlink}).
The effect of the one-loop correction is to increase $\xi$ by about $10\%$. The largest error from using $\bar{\xi}$ was found to be $\mathcal{F}^{\bar{\xi}}_{SU(2), \mathbb{BI}}\leq 8(4)\%$. In contrast, we observe $\mathcal{F}_{SU(2)}\leq1(4)\%$ and $\mathcal{F}_{\mathbb{BI}}=1.30(13)\%$ both consistent with 0 -- albeit the $SU(2)$ Monte Carlo results have larger uncertainties compared to the $U(1)$ case. This agreement is found in both $2+1d$ and $3+1d$ $\mathbb{BI}$ results (see \tab{gateperlink}).

\begin{table}[ht]
    \caption{Renormalized Anisotropies from 1-loop calculation, discrete group $\mathbb{BI}$, and \cite{Teper:1998te}.}
    \label{tab:gateperlink}
    \begin{tabular}{cllc|ccc}
    \hline\hline
    $\beta_{\xi}$& $N_s$ & $N_t$ & $\bar{\xi}$ &$\xi_{\rm 1-loop}$ & \multicolumn{2}{c}{$\xi$}\\\hline
    &&&&&$\mathbb{BI}$&$SU(2)$~\cite{Teper:1998te}\\
    \hline
    $D=3$&&&&&&\\
    2.00&36&72&2.00&2.097&2.099(1)&$\cdots$\\
2.00&12\footnote{\label{tfv}This is the largest volume simulated}&60\footref{tfv}&4.00&4.278&$\cdots$&4.35(19)\\
2.65&16\footref{tfv}&64\footref{tfv}&4.00&4.207&$\cdots$&4.22(11)\\
3.00&36&72&1.33&1.351&1.369(19)&$\cdots$\\
    4.00&24\footref{tfv}&96\footref{tfv}&4.00&4.136&$\cdots$&4.08(9)\\
    \hline
    $D=4$&&&&&&\\
    3.0&36&72&1.33&1.351&1.36(1)&$\cdots$\\
    \hline
    \end{tabular}
\end{table}

In all the groups studied, $\mathcal{F}_g$ was found to decrease or remain constant as $\beta_{\xi}$ was increased -- in agreement with expectation for a weak-coupling calculation, with the caveat that for discrete groups, $\beta_{\xi}$ should be away from the freezing-out regime.
For all the $\beta_{\xi}$ values investigated here, we obtain that the systematic error of approximating the one loop results to the non-perturbative results for both discrete group and continuous group is less than 10\%. Given that these $\beta_{\xi}$ values are corresponding to relative strong coupling, 10\% systematic error are conservative for realistic simulations using quantum computers where larger $\beta_{\xi}$ values are used.

\section{\label{sec:Conclusions}Conclusions}
Quantum field theories simulated with quantum computers are naturally lattice-regularized theories, requiring renormalization before comparisons to
experiments can be made. Quantum simulations are constructed within the Hamiltonian formalism, where a spatial lattice with spacing $a$ is
time-evolved. Further approximations are required, as the time evolution operator $\mathcal{U}(t)$ built from the Kogut-Susskind Hamiltonian usually cannot be exactly implemented in an efficient manner. One common method for these approximations is trotterization, which introduces finite temporal lattice spacings $a_t$ and thus a finite anisotropy factor $a/a_t$ in the quantum simulations. As the trotterized $\mathcal{U}(t)$ can be related to the Euclidean transfer matrix on the anisotropic lattice via analytical continuation,  it is thus beneficial to have the perturbative relations between the bare and renormalized quantities in Euclidean spacetime, e.g. the anisotropy factor $\xi$ as a function of $\beta_\xi$ and $\bar{\xi})$. 

In the near term, studies of quantum field theory on quantum computing will be limited to low-dimensions at coarse $a_t$. In this article, we extended the perturbative matching of coupling constants to general $SU(N)$ and $U(N)$ gauge theories for any anisotropy factor $\xi$ and general dimensions. The results presented here can be easily used for Euclidean measurements as well as inputs to quantum simulations through analytical continuation. As examples, we compared anisotropy factors obtained via the one-loop Renormalization to those determined from Monte Carlo simulations,
and found great agreement for $SU(N)$ gauge theories. For $U(1)$ gauge theories, the one-loop calculation corrects most of the renormalization effects observed in the non-perturbative results. To best of our knowledge, these comparisons were not previously performed before and provide important guidance for the validity of the perturbative calculations. Taken holistically, our results suggest that the one-loop $\xi$ can serve as a replacement for the nonperturbative value in lattice calculations while inducing a systematic error $\leq10\%$, with $SU(2)$ appearing to have better agreement than $U(1)$ in $2+1d$. In the weak coupling regime at sufficiently small $a$, this error is subleading to quantum errors for near term quantum simulations. Comparing the $\xi$ parameters calculated perturbatively for continuous groups with those calculated non-perturbatively for discrete groups, we find satisfactory agreement, suggesting that the perturbative relations derived in this paper are also applicable to discrete groups in the parameter space of interest.

\begin{acknowledgments}
This work is supported by the Department of Energy through the Fermilab QuantiSED program in the area of ``Intersections of QIS and Theoretical Particle Physics". Fermilab is operated by Fermi Research Alliance, LLC under contract number DE-AC02-07CH11359 with the United States Department of Energy. 
\end{acknowledgments}

\appendix
\begin{widetext}
\section{Important Integrals for Effective Action}
\label{sec:append}
In the course of deriving the effective action, a number of integrals are obtained that need to be evaluated numerically.  We have collated them here. Using the abbreviation:
\begin{eqnarray}
    b^2 = \sum^{D-1}_{i=1} \sin^2 x_i,
\end{eqnarray}
we have
\begin{equation}\label{eq:I_integrals}
I_1(\xi) = \xi \bigg(\frac{2}{\pi}\bigg)^{D-1}\int^{\pi/2}_0 d^{D-1}x ~\frac{b}{ \sqrt{\xi^2 + b^2}}\end{equation}
\begin{equation}
I_{2a}(\xi)=\xi\bigg(\frac{2}{\pi}\bigg)^{D-1}\int^{\pi/2}_0 d^{D-1}x~ \frac{\xi^2+ 2b^2}{b(\xi^2+ b^2)^{3/2}}\end{equation}
\begin{equation}
I_{2b}(\xi)= \xi^3\bigg(\frac{2}{\pi}\bigg)^{D-1}\int^{\pi/2}_0 d^{D-1}x~ \frac{1}{b\sqrt{\xi^2+ b^2}(b+\sqrt{\xi^2+ b^2})} \end{equation}
\begin{equation}
I_3(\xi)= \xi\bigg(\frac{2}{\pi}\bigg)^{D-1}\int^{\pi/2}_0 d^{D-1}x~ \frac{\sin^2 x_1\sin^2 x_2 (\xi^2+ 2b^2)}{b^{3}(\xi^2+ b^2)^{3/2}} \end{equation}
\begin{equation}
I_4(\xi)= \xi\bigg(\frac{2}{\pi}\bigg)^{D-1}\int^{\pi/2}_0 d^{D-1}x~ \frac{\sin^2 2x_1 (\xi^2+ 2b^2)}{b^{3}(\xi^2+ b^2)^{-3/2}}\end{equation}
\begin{equation}
I_5(\xi) = \xi^2 \bigg(\frac{2}{\pi}\bigg)^{D-1}\int^{\pi/2}_0 d^{D-1}x~\frac{1}{(\sqrt{\xi^2 + b^2})(b+ \sqrt{\xi^2 + b^2})}\end{equation}
\begin{equation}
I_6(\xi) =\xi^3\bigg(\frac{2}{\pi}\bigg)^{D-1}\int^{\pi/2}_0 d^{D-1}x~ \frac{\cos^2 x_1}{b(\xi^2+ b^2)^{3/2}}\end{equation}
\begin{equation}
\label{eq:i_last}
I_7(\xi)=\xi^3\bigg(\frac{2}{\pi}\bigg)^{D-1}\int^{\pi/2}_0 d^{D-1}x~
\frac{1}{b(\xi^2+ b^2)^{3/2}}
\end{equation}

\section{Series expansion of \texorpdfstring{$I_1(\xi)$}{I1(x)} in terms of \texorpdfstring{$\xi^{-2n}$}{x-2n}}
\label{apx:series}
For the future study of the renormalization of the Minkowski spacetime anisotropy via analytical continuation, it's useful to obtain the series expansion of the integrals Eqs.~(\ref{eq:I_integrals})-(\ref{eq:i_last}) in terms of $\xi^{-1}$. In this appendix, we will study $I_1(\xi)$ as an example and give the special functions related to its expansion. Expanding $(1+b^2/\xi^2)^{-1/2}$, $I_1(\xi)$ can be written as
\begin{align}
    I_1(\xi) = \bigg(\frac{2}{\pi}\bigg)^{D-1}\int^{\pi/2}_0 d^{D-1}x \sum_{k=0}^\infty \frac{\Gamma(1/2)}{\Gamma(1/2-k)k!} \frac{b^{2k+1}}{\xi^{2k}}= \sum_{k=0}^\infty \frac{\Gamma(1/2)}{\Gamma(1/2-k)k!\xi^{2k}} \braket{b^{2k+1}}_{D-1},
\end{align}
where we have defined
\begin{align}
     \braket{b^{2k+1}}_{D-1}\equiv \bigg(\frac{2}{\pi}\bigg)^{D-1}\int^{\pi/2}_0 d^{D-1}x \; b^{2k+1}.
\end{align}

To evaluate $\braket{b^{2k+1}}_{D-1}$, we define the distribution function $g_{D-1}(b^2)$:
\begin{align}
    g_{D-1}(b^2)\equiv \bigg(\frac{2}{\pi}\bigg)^{D-1}\int^{\pi/2}_0 d^{D-1}x \;\delta\left(\sum_{i=1}^{D-1} \sin^2 x_i-b^2\right)
\end{align}
and thus 
\begin{align}
  \braket{b^{2k+1}}_{D-1}=\int_0^{D-1}db^2 g_{D-1}(b^2)b^{2k+1}
\end{align}
The Fourier transform of $g_{D-1}$ is
\begin{align}
    \mathcal{F}\{g_{D-1}\}(\omega)&=\int_0^{D-1}db^2 e^{-i\omega b^2}\bigg(\frac{2}{\pi}\bigg)^{D-1}\int^{\pi/2}_0 d^{D-1}x \;\delta\left(\sum_{i=1}^{D-1} \sin^2 x_i-b^2\right)\notag\\
    &=\prod_{i=1}^{D-1} \left [\frac 2 \pi \int^{\pi/2}_0 dx_i \exp(-i \omega \sin^2 x_i)  \right]\notag\\
    &=\left[e^{-i\omega/2} J_0 (\omega/2)\right]^{D-1}
\end{align}
Where $J_0(x)$ is the Bessel function of the first kind of order zero. The inverse Fourier transform has a simple analytic expression for $D-1=1,2$:
\begin{align}\label{eq:g1}
    g_1(b^2)=\frac{1}{\pi \sqrt{b^2(1-b^2)}}, \quad 0<b^2<1,
\end{align}
\begin{align}\label{eq:g2}
     g_2(b^2)=\frac{2}{\pi^2} K[ 1- (b^2-1)^2],\quad 0<b^2<2.
\end{align}
Where $K(k)=F(\frac{\pi}{2} | k)=\int_0^{\pi/2}d\theta(1-k\sin^2\theta)^{-1/2}$ is the incomplete elliptic integral of the first kind with the upper limit specified. For higher dimensions, we can use the following relation between $g_{D-1}$ and $g_{D}$:
\begin{align}\label{eq:gDconv}
    g_{D}(b^2)=\frac{2}{\pi}\int_0^{\pi/2} dx_D g_{D-1}(b^2-\sin^2 x_D)=\int_0^1 du g_1(u)g_{D-1}(b^2-u).
\end{align}
\begin{table}[]
    \centering
    \begin{tabular}{c|cccc|cc}
    \hline\hline
        $D-1$ & $\braket{b}_{D-1}$ &$\braket{b^3}_{D-1}$ & $\braket{b^5}_{D-1} $& $ \braket{b^7}_{D-1}$ & $\braket{b}_{\mathrm{Gaussian}}$ & $\sqrt{(D-1)/2-1/16}$\\
    \hline
         1 & 0.63662 & 0.424413 & 0.339531 & 0.291026 &0.677765 &0.661438\\
         2 & 0.958091 & 1.09818 & 1.46262 & 2.13298 &0.969799 & 0.968246\\
         3 & 1.1938 & 1.9557 & 3.60865 & 7.22728 &1.19719 & 1.19896\\
    \hline
    \end{tabular}
    \caption{$\braket{b^{2k+1}}_{D-1}$ values for $D-1=1,2,3$. Columns 2-5 are computed from the exact distribution functions \eq{g1}, \eq{g2} and \eq{gDconv}, while the last two columns are from \eq{gDGaussian} and \eq{b_est} respectively.}
    \label{tab:b_averages}
\end{table}
For $D-1=1,2,3$, the lowest few $\braket{b^{2k+1}}$ are listed in \tab{b_averages}. Noticing that $\braket{b}_{D-1}$ determines the dimensional dependence of the limit $I_1(\xi\to \infty)$, it's helpful to derive an analytical estimate of $\braket{b}$ in higher dimensions. As each $\sin x_i^2$ has a mean value of $1/2$ and a variance of $1/8$ independently, for a large $D$, $g_{D-1}(b^2)$ can be approximated as the Gaussian distribution $\mathcal{N}(\frac{D-1}{2},\frac{D-1}{8})$, with the normalization adjusted to its range $[0,D-1]$:
\begin{equation}\label{eq:gDGaussian}
    g_{D-1}(b^2)\approx 
        \exp \left [ -\frac{4(b^2-D+1)^2}{D-1} \right]\frac{2}{\sqrt{\pi (D-1)}\mathrm{Erf}(\sqrt{D-1})}, \quad 0<b^2<D-1
\end{equation}
Approximating the variance of $b$, $\sigma_b^2$ as 
\begin{align}
    \sigma^2_b\approx\sigma^2_{b^2}\times\left(\frac{db}{db^2}\right)^2  \bigg|_{b^2=(D-1)/2}= \frac{D-1}{8}\times\left(\frac{1}{\sqrt{2(D-1)}}\right)^2=\frac{1}{16},
\end{align}
We get
\begin{align}\label{eq:b_est}
    \braket{b}_{D-1}= \sqrt{\braket{b^2}_{D-1}-\sigma^2_b}\approx\sqrt{\frac{D-1}{2}-\frac{1}{16}}. 
\end{align}
Both \eq{gDGaussian} and \eq{b_est} give good approximation for $\braket{b}_{D-1}$ for $D-1=2,3$, as listed in \tab{b_averages}. Using \eq{b_est}, the large anisotropy limit in higher dimensions reads,
\begin{align}
    I_1(\xi\to \infty) =\braket{b}_{D-1} \approx \sqrt{\frac{D-1}{2}-\frac{1}{16}}.
\end{align}
\end{widetext}

\bibliography{refs}

\end{document}